\documentclass[a4paper,12pt]{article}
\usepackage[cp1250]{inputenc}
\usepackage{epsfig}
\usepackage{amssymb}
\usepackage[english]{babel}
\usepackage{amsmath}
\usepackage{color}
 0

\title{Quantum corrections to $\phi^4$ model solutions and applications to Heisenberg chain dynamics}
\author{Grzegorz Kwiatkowski and Sergey Leble,\\ leble@mifgate.pg.gda.pl,  gkwiatkowski@mifgate.pg.gda.pl\\
  Gda\'nsk University of Technology, \\
ul. G. Narutowicza 11/12, 80-952 Gda\'nsk, Poland\\}
\begin{document}
 \maketitle

 \begin{abstract}
 The Heisenberg spin chain is considered in $\phi^4$ model approximation.
 Quantum corrections to  classical solutions of the one-dimensional
$\phi^4$ model within the correspondent physics are evaluated with account of rest $d-1$ dimensions of a d-dimensional theory.
A quantization of the models is considered in terms of space-time functional integral. The generalized zeta-function formalism is used 
to renormalize and evaluate the functional integral and quantum corrections to energy in quasiclassical
approximation. The results are applied to appropriate conditions of the spin chain models and its dynamics, 
which elementary solutions, energy and  the quantum corrections are   calculated.
\end{abstract}

\section{Introduction}
 There is a wide field of 
 Heisenberg spin chain \cite{Heis} realizations, intensely studied as quantum integrable system \cite{M} and in the context of its static and  dynamic properties in external magnetic field \cite{M,jedn}.
 The original approach  of W. Heisenberg is based on a localized electrons as initial approximation, valid for metals with weak conductance.  The second approximation accounts quantum exchange (due to Pauli principle) between electrons in different places. Starting from Heitler-London formula for the exchange and Coulomb integrals, one arrives at a (Heisenberg) Hamiltonian which, by construction, describes spin system field of a solid. Investigations of symmetry in the first paper of Heisenberg \cite{Heis} gives a fundamental approach to magnetics classification as well as to ferromagnetism phenomenon and, for example,  its existence only in cubic crystals (eight neighbors necessity). All this allows to believe in further 
 developmenmt as the whole model as its particular cases applications.

 Some results of the known model applications are  directly related to experiment in thermodynamics context, e.g. 	in \cite{YKSM}. Easy-plane ferromagnetism and in-plane domain-wall form factor is theoretically studied in connection with neutron scattering effects in $CsNiF_3$ crystal \cite{Fog}.
Let us stress its  quantum origin, which apart from well-known linear quasiparticles, provides a way to account for nonlinear collective phenomena as  kinks, solitons and cnoidal waves. Generally speaking we have important applications of  soliton theory  aspects, as, e.g. in  \cite{Mikeska}. Some papers reduce the theory to  Sine-Gordon model \cite{KL}. Other option leads to non-integrable $\phi^4$ model, which solutions are very similar. We would note a growing interest to a secondary quantization of such nonlinear quasiparticle fields,  that, for example, allows to obtain so-called quantum corrections to classical energy of the objects.

 There is a method that is convenient for evaluation of the corrections. It is Feynmann  integral by trajectories \cite{F} (continual integral). The quantum corrections is a topic of stable interest since the seminal paper  of R.F. Dashen, B. Hasslacher and A. Neveu \cite{DHN}, see also L. D. Faddeev, L.A. Takhtajan and V. E. Korepin papers \cite{FT}.
The functional integral method becomes a practical tool for evaluation of quasiclassical corrections to the action from the times of V.P. Maslov paper \cite{Mas}.

 One of principal results of the method is obtained in \cite{Bor0}, where the general algorythm of corrections evaluation is elaborated for arbitrary background profiles, expressions for ground state energies were derived for a 3+1 dimensions theory with a potential dependant on a single variable. Generalization for the supersymmetric kink is given in \cite{Bor}. Solutions for Sine-Gordon quasiperiodic potentials was given in \cite{Paw}.

 Developing these results to arbitrary dimensions, investigating  kink models and periodic solutions we demonstrated details of the Feynmann integral construction and generalized zeta function evaluation as well as the renormalization realization \cite{KL,KLH}.  A general algebraic method of quantum corrections evaluation based on zeta-function \cite{Kon} is used and the Green function for heat equation with an elliptic potential is constructed (see also \cite{L}).

 In this paper we continue our investigations of the problem in the spirit of \cite{KL} and fix our attention on Heisenberg chain model  in  an  so-called $\phi^4$ or Landau-Ginsburg \cite{GL} approximation (see also Gross-Pitaevski equation \cite{GP}) approximation. The merit of the presented paper is the attempt to derive conditions of possible application to realistic conditions of a magnetic medium (Sec. 2).
In the next section we describe general features of the Heisenberg spin chain model and its reduction in specific conditions of $\phi^4$. In Sec. 3 we reproduce formulas resulting form \cite{KL} for readers convenience, namely, the space-time consideration, close to original Feynmann papers \cite{F}. 
As for terminology, we use the word renormalization \cite{KL} to exclude divergence terms while in some other papers the term regularisation is met.
The final section is devoted to specific case of so-called zero "mass" $m^2= 0$ condition, that is specified by a distinguished value of magnetic field, in which the
field configuration  drastically changes.

\section{The $\phi^4$ model of Heisenberg spin chain}

\subsection{General equation of motion}
According to \cite{jedn} (with $\overrightarrow{S}_n=(S_n^x,S_n^y,S_n^z)$ as unit vectors) the Heisenberg magnetic chain with anisotropy in the direction of the chain and external magnetic field perpendicular to the chain is described by a classical Hamiltonian
\begin{equation}
	H=-J\sum_n \overrightarrow{S}_n\cdot \overrightarrow{S}_{n+1}+D\sum_n (S_n^z)^2 -g\mu_B B \sum_n S_n^x
\end{equation}
with corresponding equation of motion (here in SI units)
\begin{equation}
	\hbar \partial_t \overrightarrow{S}_n = \overrightarrow{S}_n\times (-J (\overrightarrow{S}_{n+1}+ \overrightarrow{S}_{n-1})+ 2DS_n^z \hat{z} -g\mu_B B  \hat{x}),
\end{equation}
where $J$, $D$ are spin coupling constants and $g$ is the effective electron g-factor, $\mu_B$ is the Bohr magneton, $\hat{x}$ and $\hat{z}$ are versors and $B$ is the magetic field. After taking the continuum limit (with $a$ as lattice constant) one obtains
\begin{equation}
\left\{\begin{array}{c}
\hbar\partial_t S^x=-Ja^2(S^y\partial^2_z S^z-S^z\partial^2_z S^y)+2DS^y S^z \\
\hbar\partial_t S^y=-Ja^2(S^z\partial^2_z S^x-S^x\partial^2_z S^z)-2DS^x S^z -g\mu_B B S^z \\
\hbar\partial_t S^z=-Ja^2(S^x\partial^2_z S^y-S^y\partial^2_z S^x) +g\mu_B B S^y
\end{array}\right. .
\end{equation}
By substituting $\overrightarrow{S}=(\cos\theta \cos\phi,\sin\theta \cos\phi, \sin\phi)$ one can reduce equations of motion to
\begin{equation}
\begin{array}{c}
\hbar\cos\phi\partial_t \theta= Ja^2(\partial_z^2\phi+\sin\phi\cos\phi(\partial_z \theta)^2) -2D\cos\phi \sin\phi -g\mu_B B \sin\phi\cos\theta , \\
	\hbar \partial_t \phi = -Ja^2(\cos\phi\partial_z^2\theta-2\sin\phi\partial_z \theta \partial_z \phi) +g\mu_B B \sin\theta .
\end{array}
\end{equation}

\subsection{Model $\phi^4$  approximation}
Stationary points of the system for $D<0$ (easy axis anisotropy) are such pairs $(\theta,\phi)$ for which $\theta=0,\ \phi\in\{-\arccos(\frac{g\mu_B B}{-2D}),0,\arccos(\frac{g\mu_B B}{-2D})\}$ with $\phi=0$ being unstable.
For $g\mu_B B$ close to $-2D$ both stable points are around $\phi=0$. In such a situation it is valid to assume $\phi\approx 0$ (with $\phi^3$ as the highest considered term) and with $\theta$ as the highest considered term.
\begin{equation}
\begin{array}{c}
\hbar\partial_t \theta= Ja^2\partial_z^2\phi -2D(\phi-\frac{2\phi^3}{3}) -g\mu_B B(\phi-\frac{\phi^3}{6}) , \\
	\hbar \partial_t \phi = -Ja^2\partial_z^2\theta +g\mu_B B \theta .
\end{array}
\end{equation}
If we additionally assume $|Ja^2\partial_z^2\theta|<<|g\mu_B B\theta|$, we obtain
\begin{equation}
\begin{array}{c}
\hbar\partial_t \theta= Ja^2\partial_z^2\phi -2D(\phi-\frac{2\phi^3}{3}) -g\mu_B B(\phi-\frac{\phi^3}{6}) , \\
	\hbar \partial_t \phi =g\mu_B B \theta ,
\end{array}
\end{equation}
which leads to
\begin{equation}\label{E}
\begin{array}{c}
\frac{\hbar^2}{g\mu_B B}\partial_t^2 \phi= Ja^2\partial_z^2\phi -(2D+g\mu_B B)\phi+\frac{8D+g\mu_B B}{6}\phi^3 , \\
	\theta=\frac{\hbar}{g\mu_B B} \partial_t \phi.
\end{array}
\end{equation}
The result represents the $\phi^4$ model with the energy density
\begin{equation}\label{Ham}
    H=\frac{\hbar^2}{2ag\mu_b B}\left(\frac{\partial\phi}{\partial t}\right)^2+\frac{Ja}{2}\left(\frac{\partial\phi}{\partial z}\right)^2+\frac{2D+g\mu_B B}{2a}\phi^2-\frac{8D+g\mu_B B}{24a}\phi^4 .
\end{equation}
After rewriting the equations in dimensionless variables ($z=a z',\ t=T t'$ with $T$ as the time scaling parameter in the Feynman integral as in \cite{KL}) we obtain
\begin{equation}
	\left\{\begin{array}{c}
	\frac{\hbar^2}{T^2 g\mu_B B}\partial_{t'}^2 \phi= J\partial_{z'}^2\phi -(2D+g\mu_B B)\phi+\frac{8D+g\mu_B B}{6}\phi^3 \\
 H=\frac{\hbar^2}{2g\mu_b B T^2}\left(\frac{\partial\phi}{\partial t'}\right)^2+\frac{J}{2}\left(\frac{\partial\phi}{\partial z'}\right)^2+\frac{2D+g\mu_B B}{2}\phi^2-\frac{8D+g\mu_B B}{24}\phi^4
\end{array}\right. .
\end{equation}
For simplicity of further calculations we will write
\begin{equation}\label{resc}
	\left\{\begin{array}{c}
	V^2=\frac{6(2D+g\mu_B B)}{8D+g\mu_B B} \\
	m^2=-\frac{2D+g\mu_B B}{J} \\
	c^2=\frac{J g\mu_b B T^2}{\hbar^2}
	\end{array}\right. ,
\end{equation}
where $m$ and $V$ are parameters of the potential and $c$ is a dimensionless propagation speed. Energy density   takes the form
\begin{equation}
	H=\frac{J}{2}\left(\frac{1}{c^2}\left(\frac{\partial\phi}{\partial t'}\right)^2+\left(\frac{\partial\phi}{\partial z'}\right)^2-m^2\phi^2+\frac{m^2}{V^2}\phi^4\right) .
\end{equation}
\subsection{Static kink of the  $\phi^4$ model}
For the above described system there exists a well known static kink solution
\begin{equation}
\phi=V\tanh\left(m z'\right),
\end{equation}
which in this case represents a crossection of a flat, uniform domain wall in direction of it's normal vector. Classical energy of the kink is given by integration of the energy density (\ref{Ham})
\begin{equation}
	E_c=\frac{11 J mV^2}{12}=\frac{11\sqrt{-J}(2D+g\mu_B B)^{\frac{3}{2}}}{2\sqrt{2}(8D+g\mu_B B)}.
\end{equation}
For a domain wall this represents the energy per single chain of atoms. It is of note, that kink solutions vanish, when $g\mu_B B$ reaches $-2D$. There are however static solutions, which are still present for $g\mu_B B\geq -2D$ and an example will be discussed in section (\ref{m0}).
\section{Quasiclassical Quantum Corrections}

\subsection{Quantization Scheme}
For the purpose of this publication we will use a semi-classical quantization procedure explained in detail in \cite{KL}. For a given classical system described by action integral $S$ with a static solution $\phi$ we derive energy corrections by expanding the action in path integral formulation of the propagator
\begin{equation}
    \langle\phi|e^{-\frac{i}{\hbar}TH}|\phi\rangle=\int_{C^{0,T}_{\phi,\phi}}D\varphi(x,t)e^{\frac{i}{\hbar}S(\varphi)}
\end{equation}
in a Taylor series around the classical solution and cutting it at the first non-trivial term. Then the formal expression for quantum correction to energy is
\begin{equation}
	\Delta E = -\frac{\hbar}{iT}\ln\left(\det\left[L\right]\right),
\end{equation}
where $L$ is the second derivative of the classical Lagrangian up to a multiplicative constant arising from the gaussian integrals during the derivation process (see \cite{Mas} and \cite{KL}). We use zeta-function renormalization scheme to deal with emerging infinities
\begin{equation}\label{energy}
	\Delta E= -\frac{\hbar}{iT}\lim_{s\rightarrow 0_+}\frac{\partial}{\partial s}\int_0^{\infty}\tau^{s-1} \int(g_L(\tau,\overrightarrow{x},\overrightarrow{x}) -g_{L_0}(\tau,\overrightarrow{x},\overrightarrow{x}))d\overrightarrow{x}d\tau,
\end{equation}
where $\overrightarrow{x}$ covers all variables of the classical system, $g_L$ is the Green function of the heat equation
\begin{equation}\label{gd}
	\left(\frac{\partial}{\partial \tau}+L\right)g_L(\tau,\overrightarrow{x},\overrightarrow{x}_0)=\delta(\tau)\delta(\overrightarrow{x}-\overrightarrow{x}_0)
\end{equation}
and $L_0$ is an operator analogous to $L$ with a constant potential (representation of vacuum). Additionally, mass scale is used to cut logarithmic divergence in all relevant parameters (see \cite{Kon} and \cite{KL}). Often the intermediate steps of (\ref{energy}) are defined explicitly as
\begin{equation}\label{gamma}
	\gamma(\tau)=\int(g_L(\tau,\overrightarrow{x},\overrightarrow{x})-g_{L_0}(\tau,\overrightarrow{x},\overrightarrow{x}))d\overrightarrow{x}
\end{equation}
and
\begin{equation}
	\zeta(s)=\frac{1}{\Gamma(s)}\int_0^{\infty}\tau^{s-1}\gamma(\tau)d\tau.
\end{equation}
It was shown in \cite{L}, with important link to generalized zeta function theory, that if $L$ can be written as a sum of operators acting on independent variables $L=\sum_i L_i$, heat equation Green function for $L$ can be written as a product of Green functions for $L_i$. We are using this property to account for arbitrary number of spatial variables of the classical system.
\subsection{Quantum Corrections to $\phi^4$ kinks}
We now procede to calculate quantum corrections for energy using the above described generalized zeta function renormalization scheme with following form of operators:
\begin{equation}
	L_1=A\left(\frac{\partial^2}{\partial z'^2}-4m^2+6m^2 sech^2(m z')\right) ,
\end{equation}
\begin{equation}
	L_2=-\frac{A}{c^2}\frac{\partial^2}{\partial t'^2} ,
\end{equation}
\begin{equation}
	L_3=\frac{A}{l^{d-1}}\Delta_{d-1} ,
\end{equation}
\begin{equation}\label{A}
	A=\frac{iTJ}{2\pi\hbar r^2} ,
\end{equation}
where $d$ is the total number of spatial dimensions, $\Delta_{d-1}$ covers all spatial variables except for $z'$, $l$ is the range of all additional spatial dimensions in multiple of $a$ (due to the same rescaling as for $z$) and $r$ is the mass scale. Laplace transform of the Green function diagonal for $L_1$ was derived by use of algorithm described in \cite{KLH}. For $L_2$ and $L_3$ spectra continuum approximation was taken. We obtained following corrections
\begin{equation}
	\Delta E_{d=1}=\frac{\hbar cm} {2T\pi}\left(2+\frac{\pi}{\sqrt{3}}-2\ln(2) -3\ln(-Am^2)\right),
\end{equation}
\begin{equation}
	\Delta E_{d=2}=-\frac{\hbar cm^2 l} {2T\pi} \left(3+\frac{3}{2}\arcsin\left(\frac{1}{\sqrt{3}}\right)\right),
\end{equation}
\begin{equation}
	\Delta E_{d=3}=-\frac{\hbar cm^3 l^2} {8T\pi^2} \left( -6\ln(-Am^2)-6 +\frac{2}{9}(-11+3\sqrt{3}\pi+6\ln(2))\right).
\end{equation}
Mass scale $r$ is chosen so that any logarithmic contributions will vanish ($\ln(-Am^2)=0$) for arbitrary T. Energy corrections exhibit a similar dependance on classic equation parameters as those of Sine-Gordon system \cite{KL}. Physical meaning of those parameters is however different. It is of note, that similar results were obtained by Konoplich in \cite{Kon}. Difference comes from accounting for $\left(\frac{\partial \phi}{\partial t}\right)^2$ term in classical action. After inserting proper forms of $c$ and $m$ we obtain
\begin{equation}
	\Delta E_{d=1}=\frac{\sqrt{-g\mu_B B(2D+g\mu_B B)}} {\sqrt{2}\pi}\left(1+\frac{\pi}{2\sqrt{3}}-\ln(2)\right),
\end{equation}
\begin{equation}
	\Delta E_{d=2}=\frac{\sqrt{g\mu_B B}(2D+g\mu_B B) l} {4\sqrt{J}\pi} \left(3+\frac{3}{2}\arcsin\left(\frac{1}{\sqrt{3}}\right)\right),
\end{equation}
\begin{equation}
	\Delta E_{d=3}=\frac{\sqrt{-g\mu_B B}(2D+g\mu_B B)^{\frac{3}{2}} l^2} {8\sqrt{2}J\pi^2} \left(-\frac{38}{9}+\frac{\pi}{\sqrt{3}}+\frac{2}{3}\ln(2)\right).
\end{equation}
In this system the difference between one-dimensional model and one accounting for two additional spatial dimension is especially visible near the $g\mu_B B=-2D$ border case. Since $\Delta E_{d=1}$ is proportional to $(2D+g\mu_B B)^{\frac{1}{2}}$ instead of $(2D+g\mu_B B)^{\frac{3}{2}}$, it would outweight the classical energy significantly. If we look at the ratio of corrections to classical energy for $d=3$
\begin{equation}
	\frac{\Delta E_{d=3}}{E_c}=\frac{\sqrt{g\mu_B B}(8D+g\mu_B B)}{44\pi^2 J^{\frac{3}{2}}}\left(-\frac{38}{9}+\frac{\pi}{\sqrt{3}}+\frac{2}{3}\ln(2)\right)
\end{equation}
we can see a particularly strong dependance on $J$ parameter, which represents interaction strength between neighboring electrons. Calculated ratio will be the highest for $g\mu_B B\rightarrow -2D$. In this limit we obtain
\begin{equation}
	\frac{\Delta E_{d=3}}{E_c}\propto -\left(\frac{-D}{J}\right)^{\frac{3}{2}}.
\end{equation}
It is worth noticing, that due to the way Planck constant enters propagation speed $c$, energy correction are independant of it's value. Regardless of the number of spatials dimensions taken into account, energy corrections still vanish along with the classical field, when $g\mu_B B=-2D$ (see the next section).

\section{The special case of  $2D+g\mu_B B=0$}\label{m0}
In this section we will discuss a static solution of $\phi^4$ model, which does not vanish, when both stationary points of the potential coincide. Let us rewrite equation of motion and Hamiltonian in such case, when $m^2=2D+g\mu_B B=0$ .
\begin{equation}
	\left\{\begin{array}{c}
	\frac{J}{c^2}\partial_{t'}^2 \phi= J\partial_{z'}^2\phi +D\phi^3 \\
 H=\frac{J}{2}\left(\frac{1}{c^2}\left(\frac{\partial\phi}{\partial t'}\right)^2+\left(\frac{\partial\phi}{\partial z'}\right)^2-\frac{D}{2J}\phi^4\right)
\end{array}\right. .
 \end{equation}
This system has an interesting traveling wave solution
\begin{equation}
	\phi(b(z'\pm vt'))=b\sqrt{\frac{2J(c^2-v^2)}{-D c^2}}sn(b(z'\pm vt');i)
\end{equation}
with $b$ as a free parameter real-valued amplitudes for $v<c$ due to $D<0$ ($sn$ denotes Jacobi SN elliptic function). For now we will focus on the $v=0$ case
\begin{equation}
	\phi(z')=b\sqrt{\frac{2J}{-D}}sn(b z';i).
\end{equation}
It is particularly interesting, since it's a non-trivial static solution in a system with a single stationary point $\phi=0$.
Quantum corrections to the energy of this solution can be calculated through the same procedure as for a $\phi^4$ kink with
\begin{equation}
	L_1=A\left(\frac{\partial^2}{\partial z'^2}+6b^2 sn^2(b z')\right),
\end{equation}
where $A$ is the same as in (\ref{A}) and other $L_i$ as before. Let us reproduce some formulas from \cite{L}. Laplace transform of the Green function diagonal can be obtained in the same way as for $\phi^4$ or Sine-Gordon kinks \cite{KLH}
\begin{equation}
	G_1(p,z)=\frac{p^2 - 3b^2 p (1 - z) + 9b^4 (z-2) z}{2\sqrt{(3 b^2 + p) (3 b^2 - p) p (p^2-12 b^4)}}.
\end{equation}
It is important to note, that for calculation convenience, we use rescaling $\tau\rightarrow \frac{\tau '}{A}$ for the Green function equation (\ref{gd}). The polynomial in the denominator has five simple roots. We will set the vacuum counterpart to coincide with the highest one
\begin{equation}
	G_0(p,z)=\frac{1}{2\sqrt{2b^2\sqrt{3}-p}}.
\end{equation}
We can now integrate the Green function diagonal over the period $\frac{4K(i)}{b}$, where $K$ is the complete elliptic integral of the first kind
\begin{equation}
	\hat{\gamma}(p)=\int_0^{\frac{4K(i)}{b}}(G_1(p,cn^2(b x))-G_0(p,cn^2(b x)))dx,
\end{equation}
\begin{equation}\label{hatgammap'}
\begin{array}{c}
 \hat{\gamma}(p)=\frac{6b^{4}K \left(  i\right)  +2p^{2}K\left(  i\right)  + 36b^{4}\left(  K \left(
i\right)  -E\left(  i\right)  \right)  -    3b^{2}p\left(
E\left(
i\right)-3K\left(  i\right) \right) }{b\sqrt{(3 b^2 + p) (3 b^2 - p) p (p^2-12 b^4)}} \\ -\frac{2K(i)}{b\sqrt{2b^2\sqrt{3}-p}}
\end{array}.
\end{equation}
To properly define the inverse Laplace transform, we need the $\hat{\gamma}$ function to be smooth in a $o_l<\Re(p)<o_p$ area with $o_l<o_p$ as constants \cite{Ef}. Since the $\hat{\gamma}$ function has five distinct singularities (all on the real axis), we have six potential inverse Laplace transform (disregarding the choice of the sign for the square roots), but only one of them fulfills the necessary condition $\forall_{\tau<0}\ \gamma(\tau)=0$ arising from the definition (\ref{gd}) - it occures, when $c_p\rightarrow \infty$. Thus the $\gamma$ function as defined in (\ref{gamma}) will take form
\begin{equation}
	\gamma(\tau)=\int_{o-i\infty}^{o+i\infty}e^{pA\tau}\hat{\gamma}(p)dp,
\end{equation}
where $o>2b^2\sqrt{3}$. At this point, we add all relevant variables as for $\phi^4$ case $\gamma(\tau)\rightarrow\gamma(\tau)\gamma_2(\tau)\gamma_3(\tau)$. The energy corrections per wave period are finally obtained by performing the Mellin transform and taking the derivative at $s=0$
\begin{equation}
	\Delta E=-\frac{\hbar}{iT}\lim_{s\rightarrow 0_+}\frac{\partial}{\partial s}\int_0^{\infty}\tau^{s-1}\gamma_2(\tau)\gamma_3(\tau)\int_{o-i\infty}^{o+i\infty}e^{pA\tau}\hat{\gamma}(p)dp d\tau,
\end{equation}
Plugging expressions from (\ref{hatgammap'}) and relevant forms of $\gamma_2$ and $\gamma_3$ (see \cite{KL}) yields
\begin{equation}
\begin{array}{c}
	\Delta E=-\frac{\hbar^{\frac{d}{2}}J^{\frac{1-d}{2}} \sqrt{g\mu_B B} r^d l^{d-1}}{(-2i)^{\frac{d}{2}} T^{\frac{d}{2}}}\lim_{s\rightarrow 0_+}\frac{\partial}{\partial s}\int_0^{\infty}\tau^{s-\frac{d}{2}-1}\\ \int_{o-i\infty}^{o+i\infty}e^{\frac{iTJ p\tau}{2\pi \hbar r^2}}\left(\frac{6b^{4}K \left(  i\right)  +2p^{2}K\left(  i\right)  + 36b^{4}\left(  K \left(
i\right)  -E\left(  i\right)  \right)  -    3b^{2}p\left(
E\left(
i\right) -3K\left(  i\right) \right) }{b\sqrt{(3 b^2 + p) (3 b^2 - p) p (p^2-12 b^4)}} -\frac{2K(i)}{b\sqrt{2b^2\sqrt{3}-p}}\right)dp d\tau
\end{array}.
\end{equation}
There still remains the problem of finding the explicit analitic solution of the shown integrals. Applications of the distinguished condition of the elliptic field configuration may be interesting from the point of measurements realization. Such configuration may be more easily noticed (recognized in experiments).
 \section{Conclusion}
Energy corrections for easy axis domain walls would be particularly interesting in the case of thin ferromagnetic films, where they should dominate 
the classical energy.
Formally, the results of the last section partially coincide with the known  case (see e.g. \cite{L}, where the so-called Nahm model of Yang-Mills theory is studied) but physical sense is different.

The space-time consideration we develop in our publications has obvious intensions to include mutisoliton configuration into quantum quasiclassical picture.   

\end{document}